\documentclass[
adp,fleqn]{w-art}

\usepackage{times}
\usepackage{w-thm}
\usepackage[]{graphicx}
\chardef\bslash=`\\ 

\begin{document}
\DOIsuffix{theDOIsuffix}
\Volume{12}
\Issue{1}
\Copyrightissue{01}
\Month{01}
\Year{2003}
\pagespan{1}{}
\Receiveddate{15 November 1900}
\Accepteddate{2 December 1900}




\title[Short Title]{Enhanced electron correlations in FeSb$_2$}


\author[F.\ Author]{Peijie Sun \inst{1}} 
\address[\inst{1}]{Max Planck Institute for Chemical Physics of Solids, D-01187 Dresden, Germany}
\author[S.\ Author]{Martin S\o ndergaard \inst{2}}
\address[\inst{2}]{Department of Chemistry, University of Aarhus, DK-8000 Aarhus C, Denmark}
\author[Th.\ Author]{Bo B. Iversen \inst{2}}
\author[F.\ Author]{Frank Steglich \footnote{Corresponding author.  \quad E-mail: {\sf steglich@cpfs.mpg.de}}\inst{1}}

\begin{abstract}
FeSb$_2$ has been recently identified as a new model system for studying many-body renormalizations in  a $d$-electron based narrow gap semiconducting system, strongly resembling FeSi. The 
electron-electron correlations in FeSb$_2$ manifest themselves in a wide variety of physical properties including electrical and thermal transport, optical conductivity, magnetic 
susceptibility, specific heat and so on. We review some of the properties  that form a set of experimental evidences revealing the crucial role of correlation effects in FeSb$_2$. The 
metallic state derived from slight Te doping in FeSb$_2$, which has large quasiparticle mass, will also be introduced.  
\end{abstract}
\maketitle                   





\section{Introduction}
\label{sect1}
Since the recognition of FeSi as a semiconductor with puzzling  thermodynamic properties \cite{Jacca}, the class of transition metal based, narrow gap semiconductors has been  attracting 
attention due to the presence of strong many-body renormalizations \cite{review_KI}. Interest in this class of materials was further revived by the proposal assigning FeSi to be the first 
example of a $d$-electron derived Kondo insulator (KI) \cite{fisk}, because of the strong similarities in physical properties of this compound with various $f$-electron based Kondo 
insulators.  The materials which have been argued to resemble FeSi in their physical properties include Fe$_2$VAl \cite{FeVAl}, FeGa$_3$ \cite{taka}, and FeSb$_2$ 
\cite{petrovic1,petrovic2,bentien2}. The most striking similarities to FeSi, however, were observed only in FeSb$_2$ due to their close energy scales (for example, the main transport gap in 
FeSb$_2$ and FeSi is roughly 30 and 50 meV, respectively).  Though the physical properties of FeSb$_2$ are far from being understood, an apparent consensus regarding the important role of 
electron correlations in the formation of the narrow energy gap and the relevant physical properties of FeSb$_2$, appears to have been reached. 

Different theoretical approaches have been so far proposed to capture the electron-electron correlations in this class of materials. In addition to the KI scenario, where  the small 
semiconducting gap arises in a Kondo lattice due to the hybridization of the localized $f$ (or $d$) state and a broad conduction band, Takahashi and Moriya (TM) \cite{SCR}, starting from the 
electronic structure as calculated for FeSi, have offered a different interpretation, where the unusual thermodynamics including magnetic susceptibility and specific heat can be well 
interpreted by using the spin fluctuation theory of itinerant electron systems. 
While also starting from a band insulator model, recent theoretical explorations \cite{kunes,Sentef,CBI} tend to treat FeSi and FeSb$_2$ as band insulators with strong local dynamical 
correlations, i.e., correlated band insulators (CBI) in a more general physical interpretation. 
This approach, which takes the local correlations into account via a dynamical mean-field approximation (DMFT \cite{dmft}), can also explain all major physical properties of FeSb$_2$ as well 
as FeSi \cite{kunes}. Compared to the KI scenario where an extremely narrow band width is necessary, here a more realistic band width (much larger than the gap) is employed.  In line with 
the CBI scenario, recent electronic structure calculations for FeSb$_2$ with correlation effects taken into account yield the correct, experimentally observed size of the band gaps 
\cite{Kotliar}; recently observed dispersions by angle-resolved photoemission spectroscopy (ARPES) for FeSi \cite{klein} can quantitatively be described by an itinerant behavior provided 
that an appropriate self-energy correction is included. 

FeSb$_2$ crystallizes in the marcasite-type orthorhombic structure, belonging to the space group $Pnnm$ (No. 58). While most members of this structure show semiconducting behavior, FeSb$_2$ 
appears to have the smallest energy gap and the most unusual magnetic properties \cite{marca,Fan}, 
which have been later explained by da Silver who assumed highly correlated valence and conduction bands in FeSb$_2 $\cite{silve}. Note that the recent interpretation of the magnetic 
susceptibility by the CBI model \cite{kunes} is similar to da Silver's approach. 
Electronic structure calculations for FeSb$_2$ have been performed by different authors \cite{bentien2, Kotliar, madsen, LDA}. The standard density-functional theory can yield a direct 
pseudogap of 0.2-0.3 eV  with a finite electronic density of states (DOS) in the ground state.   By including many-body effects to the electronic structure calculations through Hedin's GW 
approximation, Tomczak $et\,al.$ have recently obtained an insulating ground state for FeSb$_2$, 
with a transport gap consistent with the experimentally observed value ($\sim$ 30 meV) \cite{Kotliar}.
The similarity in the magnetic properties of FeSb$_2$ and FeSi is also seen in the calculated electronic structures: closely resembling the case of FeSi \cite{LDA_FeSi}, the local density 
approximation with on-site Coulomb repulsion correction (LDA+U) found a ferromagnetic metallic state that is nearly degenerate with the semiconducting state in FeSb$_2$ \cite{LDA}, as has 
been predicted by TM's spin fluctuation theory. Experimentally, ferromagnetism was indeed observed by doping both FeSb$_2$ and FeSi: doping the latter compound by 25\% Ge, one observes a 
first order phase transition from a paramagnetic semiconductor into a ferromagnetic metal \cite{Yeo}; doping Te \cite{Hu_Te} or Co \cite{Hu_Co} into FeSb$_2$ also results in ferromagnetic 
metallic states. Furthermore,
like FeSi can be transformed into a heavy fermion metal by Al doping \cite{FeSiAl}, FeSb$_2$ changes into a metallic state with strongly enhanced quasiparticle mass (10 $-$ 20 times of 
$m_0$) by slight Te doping \cite{sun_apl}. Note that Te substitution for Sb is non-isoelectronic and adds one electron / Te atom to the conduction band.

Except for the fundamental aspects mentioned above, our interest in this class of materials stems from their potential as thermoelectric materials in the cryogenic range. An enhanced 
thermopower has long been expected in correlated semiconductors principally due to the largely enhanced and asymmetric DOS at the Fermi level.  A colossal thermopower, ranging between $-$6 
and $-45$ mV/K at around 10 K, was observed in FeSb$_2$ \cite{bentien}. Our semi-quantitative analyses have revealed an unconventional enhancement of the thermopower, most likely due to 
many-body effects \cite{ sun_apl,sun_PRB,sun_apex}.      
In this paper, we review the typical magnetic, transport, and optical properties  as well as specific heat of FeSb$_2$ and its slightly Te doped variants. 
We will show that the classical descriptions without many-body corrections are inadequate to account for most of the observations.

\section{Magnetic properties}
\begin{figure}[htb]
\begin{minipage}[t]{.48\textwidth}
\includegraphics[width=\textwidth]{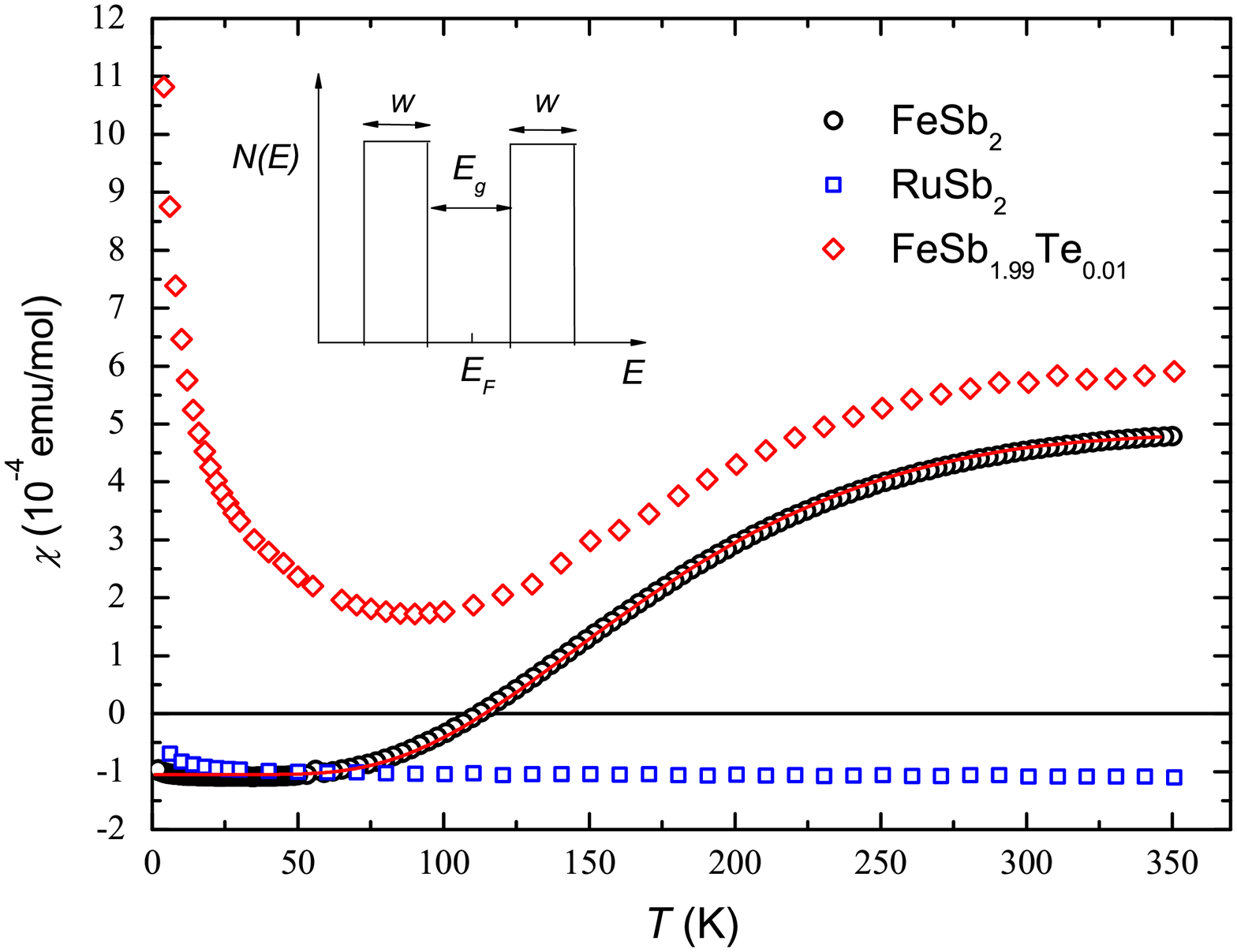}
\caption{Magnetic susceptibility $\chi(T)$ for FeSb$_2$ and a Te-doped, metallic system FeSb$_{1.99}$Te$_{0.01}$. $\chi(T)$ of RuSb$_2$ is also displayed for comparison. The solid (red) line  
is a theoretical fit (see text) of the Pauli paramagnetism for a simple narrow gap system as depicted in the inset, with a spin gap of $\Delta _s$= 76 meV and band width $W$ = 45 meV. This 
line fits the data  for FeSb$_2$ very well.}
\label{fig:chi}
\end{minipage}
\hfil
\begin{minipage}[t]{.48\textwidth}
\includegraphics[width=\textwidth]{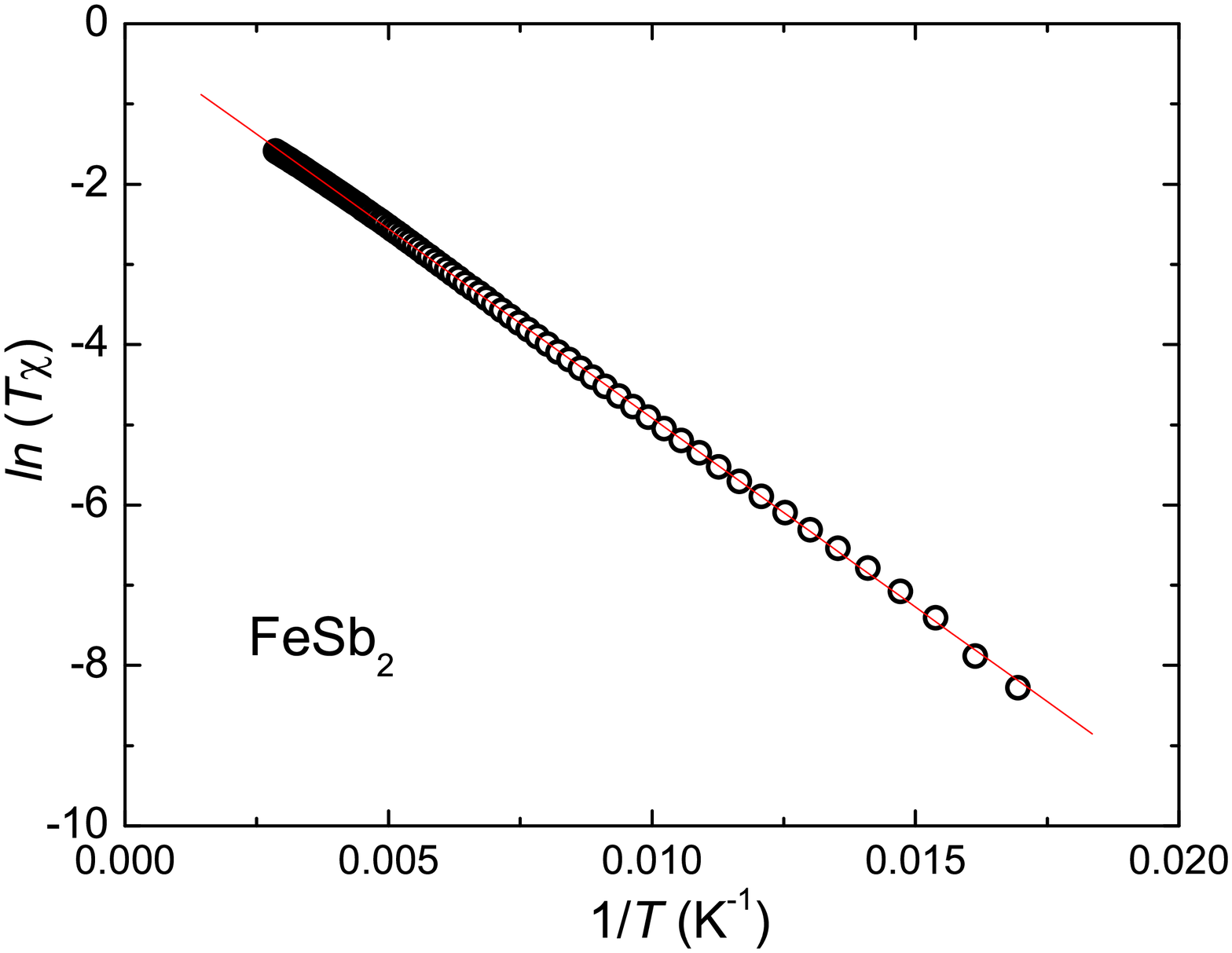}
\caption{ln($T\chi$) as a function of $1/T$ for FeSb$_2$ in the temperature range 60 and 350 K. The linear fit indicates a thermally activated CW law (see text) for describing the $\chi(T)$ 
of FeSb$_2$. A spin gap $\Delta _s =$ 81.2 meV is employed in this fit.}
\label{fig:chi2}
\end{minipage}
\end{figure}

\begin{vchtable}[htb]
\vchcaption{Different energy gaps $E_g$ derived from various measurements for FeSb$_2$ (unit: meV). While the gaps estimated from electrical transport, specific heat and quadrupole splitting 
of M\"{o}ssbauer effect are quite close to the indirect gaps estimated from optical measurements, the energy scale derived from magnetic susceptibility and nuclear spin-lattice relaxation 
time $T_1$ is much larger than the indirect gaps, and is therefore treated as a measure of the direct gap. 
However, we note that the physical interpretation of these energies could be much more complex.} 
\label{tab:1}\renewcommand{\arraystretch}{1.5}
\begin{tabular}{lllllll} \hline
  & $E_{g1}$1 (indirect) & $E_{g2}$ (indirect) & $E_{g3}$ (direct) & References\\ \hline 
Electrical resistivity   & 4$-$10 & 25$-$40 & $-$ & \cite{bentien,sun_dalton} \\
Optical conductivity   & 6  & 31 & 130 & \cite{herzog} \\ 
Magnetic susceptibility   & $-$ & $-$ & 70$-$100 & \cite{petrovic1, petrovic2,koyama,nqr1,sun_dalton} \\ 
Specific heat   & 11.2 & 50.9 & $-$ & \cite{sun_apex} \\ 
$1/T_1$ (NQR)  & $-$  & $-$ & 76.8 & \cite{nqr1,nqr2} \\ 
$^{57}$Fe M\"{o}ssbauer    & $-$ & $33$ & $-$ &  \cite{moss_FeSb2} \\ \hline
\end{tabular}
\end{vchtable}   

Shown in Fig.~\ref{fig:chi} is the magnetic susceptibility $\chi(T)$ of FeSb$_2$ and a Te-doped, metallic sample 
FeSb$_{1.99}$\-Te$_{0.01}$, together with $\chi(T)$ of the isostructural semiconductor RuSb$_2$.  
At around 100 K,  $\chi(T)$ of FeSb$_2$ exhibits a crossover from low-temperature diamagnetism to enhanced paramagnetism at higher temperatures. In contrast $\chi(T)$ of RuSb$_2$ is almost 
$T$ independent and diamagnetic due to the predominant inner-core contribution. No cooperative magnetism has been observed in FeSb$_2$ in various measurements including 
magnetic-susceptibility \cite{petrovic1,Fan,silve}, neutron-diffraction \cite{neutron_FeSb2}, and $^{57}$Fe-M\"{o}ssbauer \cite{moss_FeSb2} measurements. 
A nonmagnetic ground state of FeSb$_2$ has been commonly assumed within a simple ionic picture proposed by Goodenough \cite{goodenough}, where the iron has a low-spin 3$d^4$ configuration 
because it is located within a slightly distorted octahedron, and therefore the $t_{2g}$ state splits into two degenerate $d_{yz}$ and $d_{xz}$ orbitals (ground state, fully filled) and one 
higher lying $d_{xy}$ orbital.  This configuration, however, has not been experimentally confirmed yet. 

The thermally activated $\chi(T)$ of FeSb$_2$ above 100 K parallels that of FeSi \cite{Jacca}. Phenomenological descriptions of this unique behavior employed for FeSi include a thermally 
activated Curie law of local magnetic moments, i.e., $\chi _{cw} (T)$\,=\,$(C/T) \,{\rm exp} (-\Delta _s /k_BT)$, where $\Delta _s$ denotes the relevant spin gap. Alternatively, one may 
assume an enhanced Pauli paramagnetism of thermally activated itinerant electrons in a narrow band system (cf. inset of Fig.~\ref{fig:chi}) as obtained by an integration over the whole 
conduction band, i.e., $\chi _{pauli}(T)$ \,=\,$-2\mu _B^2 \int \! N(E) \, \partial f(E,T)/\partial E \, dE$, where $f(E,T)$ represents the Fermi function.  
The two approaches, as revealed by the calculated curves in Fig.~\ref{fig:chi2} and Fig.~\ref{fig:chi}, respectively, apply well to FeSb$_2$ and yield a similar spin gap, consistent with the 
reported values \cite{petrovic1, petrovic2,koyama,nqr1}, 70$-$100 meV. 
Note that this energy is two or three times that of the transport gap, but is close to (while somewhat smaller than) the direct gap estimated from optical conductivity (see Table\,1). 
The band width $W$ employed in the latter approach is as narrow as 45 meV, a value even smaller than the experimentally observed gap energy. Such a band width, which is unrealistic in terms 
of conventional band theory, has been taken as a proof supporting the KI scenario \cite{mandrus}.  By contrast the CBI scenario is also able to describe the observed magnetic susceptibility, 
however, with a more realistic band width (much wider than the relevant gap) \cite{kunes}.  
On the other hand, by studying the expansion coefficients of the free energy for the magnetization, Koyama $et\,al.$ \cite{koyama}  recently concluded that the magnetization process in 
FeSb$_2$ is governed by the correlated spin fluctuations, in accordance to TM's scenario.
It should be also noted that, while for FeSi, $\chi(T)$ follows a Curie-Weiss law  at temperatures above the broad maximum at around 500 K \cite{Jacca}, for FeSb$_2$, such behavior could not 
be confirmed so far.  
This is partly due to the fact that the so far available magnetic susceptibility data of the latter compound are limited to below 550 K:  FeSb$_2$ is supposed to decompose into FeSb and Sb 
above 573 K \cite{Fan}. This decomposition temperature, however, remains controversial \cite{decompose}.

Also plotted in Fig.~\ref{fig:chi} is $\chi(T)$ of a Te-doped sample, FeSb$_{1.99}$Te$_{0.01}$, which has a metallic ground state (cf. Fig.~\ref{fig:rho}). The large increase of $\chi(T)$ 
below 80 K observed for this sample could indicate the existence of local magnetic moments induced by Te. A small upturn below 20 K was also observed in some nominally pure FeSb$_2$ samples 
(see, e.g., \cite{petrovic1}). For $T$ $>$ 100 K,  the temperature dependence of $\chi(T)$ for the doped sample is very similar to that of the undoped FeSb$_2$, except for a significant 
shift upward by a constant value ($\sim$ 1.2 x 10$^{-4}$ emu/mol). Assuming this shift for FeSb$_{1.99}$Te$_{0.01}$ being due to a $T$-independent Pauli paramagnetism  derived by additional 
free charge carriers, the estimated DOS at the Fermi level is $N (\epsilon _F)$ = $\chi _{Pauli}$ / $\mu_B^2$  = 1.39 $\times$ 10$^{43}$ states$/$J mol. This value is close to $N(\epsilon 
_F)$ =  0.96 $\times$ 10$^{43}$ states$/$J mol as derived from the electronic specific heat, which is substantially enhanced by a renormalized charge carrier mass (see below).  

\section{Transport properties}
\begin{figure}[htb]
\begin{minipage}[t]{.45\textwidth}
\includegraphics[width=\textwidth]{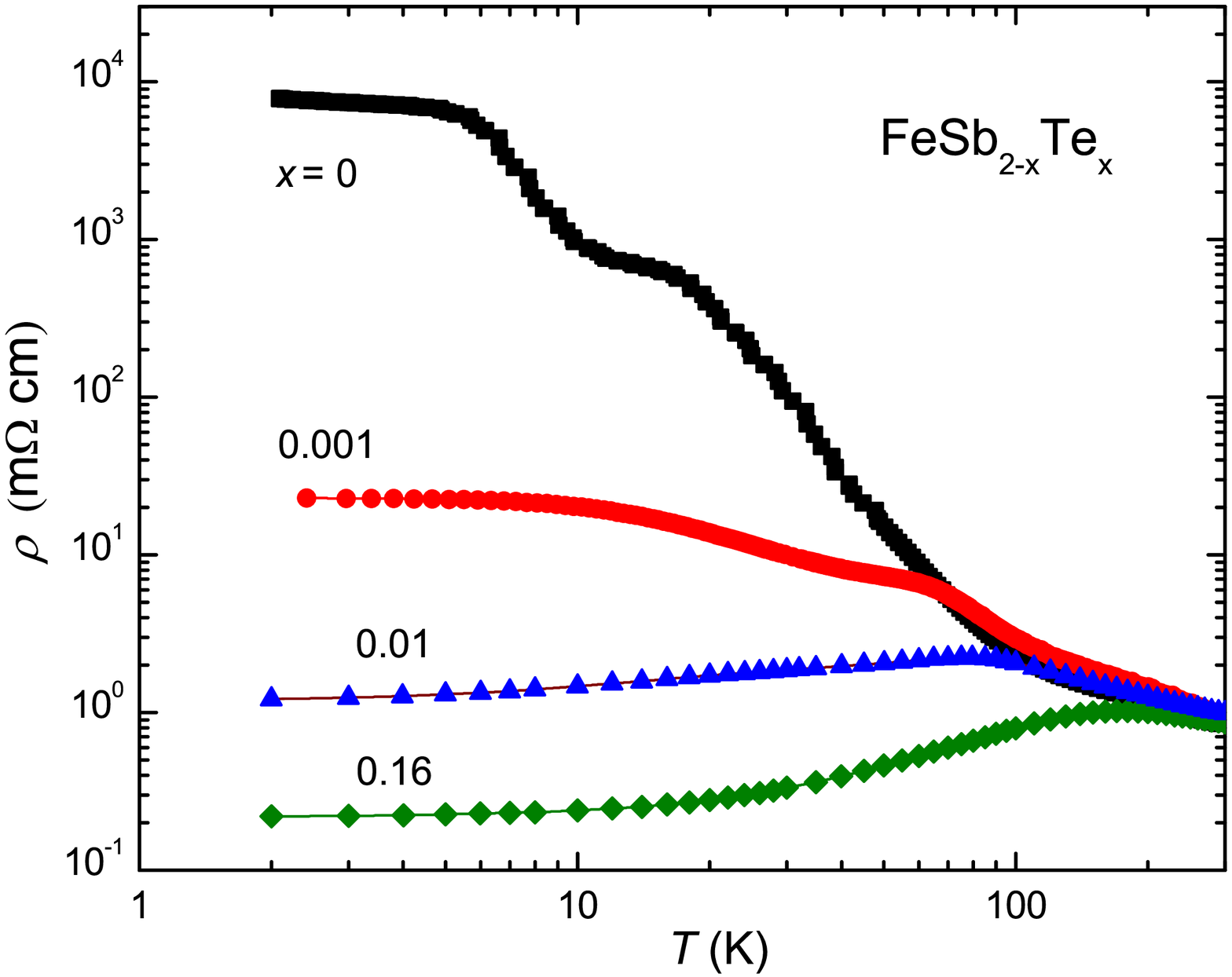}
\caption{Electrical resistivity $\rho(T)$ of FeSb$_{2-x}$Te$_x$ with varying Te content $x$\,=\,0, 0.001, 0.01 and 0.16 \cite{sun_apl}.  The former two samples are semiconductors whereas the 
two latter ones show metallic behavior below 100 K.}
\label{fig:rho}
\end{minipage}
\hfil
\begin{minipage}[t]{.45\textwidth}
\includegraphics[width=\textwidth]{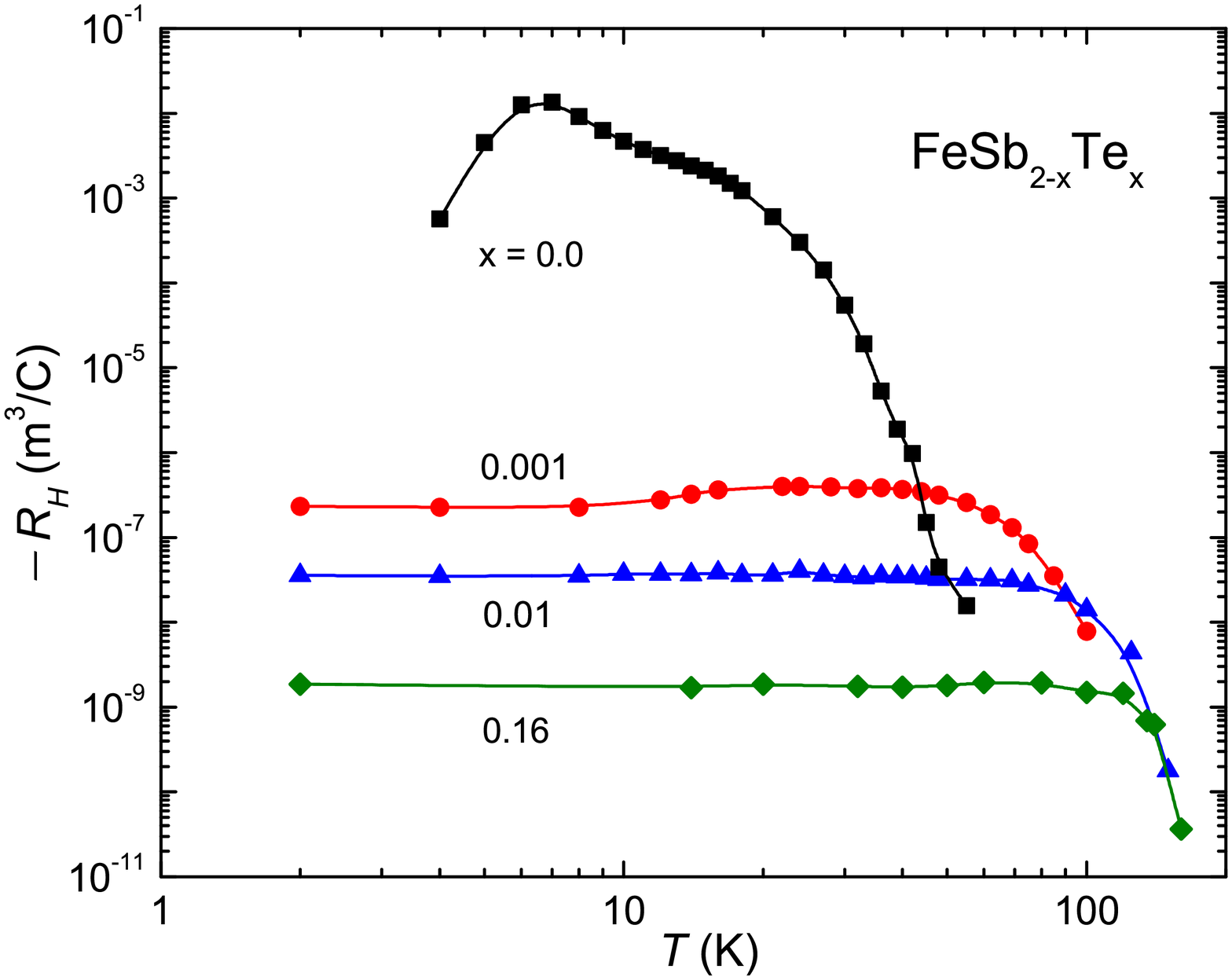}
\caption{Hall coefficient $R_H(T)$ for the same samples shown in Fig.~\ref{fig:rho}.  $R_H$ $<$ 0 in the temperature range shown for all samples indicating dominating electron transport, 
whereas at higher temperatures a sign change occurs due to competing multi-band contributions. }
\label{fig:hall}
\end{minipage}
\end{figure}

As seen in Fig.~\ref{fig:rho}, the electrical resistivity $\rho(T)$ of FeSb$_2$ is thermally activated in two temperature windows, i.e., $\rho(T)$ $\propto$ ${\rm exp} (E_g/2k_BT)$, 
separated by a shoulder at around 20\,K. Below 7 K, we observe a tendency of saturation, pointing to a residual DOS at the Fermi level. However, it is not yet clear whether this phenomenon 
is intrinsic or due to impurities.  Despite considerable variation  in the electrical resistivity from sample to sample which is typical for narrow-gap semiconductors, 
the afore-described features were found in all the single-crystalline FeSb$_2$ samples we have investigated thus far \cite{bentien,sun_dalton}.  The two transport gaps estimated for various 
samples are largely consistent, i.e., $E_{g1}$ = 4$-$10 meV and $E_{g2}$ = 25$-$40 meV for the temperature window 5$-$15 K and 50$-$200 K, respectively \cite{bentien, sun_dalton}. 

Interestingly,  a very small amount ($x$ $<$ 0.01) of Te can induce a metallic ground state in FeSb$_{2-x}$Te$_x$ (Fig.~\ref{fig:rho}). The transport properties near the semiconductor-metal 
(SM) transition were recently investigated in detail by us \cite{sun_apl}, while $\rho(T)$ over a much wider doping range ($x$ $<$ 1.2) was reported by Hu $et\,al.$ \cite{Hu_Te}.  It appears 
that, while the smaller transport gap ($E_{g1}$) is filled, the larger one ($E_{g2}$) is less influenced upon increasing the Te concentration up to at least $x$ = 0.01. This feature is 
consistent with the magnetic susceptibility of FeSb$_{2-x}$Te$_x$ (cf. Fig.~\ref{fig:chi}), which persists to be thermally activated at $T$ $>$ 100 K. The Hall coefficient $R_H$ of the 
metallic samples exhibits nearly constant values at low temperatures (cf. Fig.~\ref{fig:hall}), which suggests that each Te atom introduces one extra electron to the conduction band 
\cite{sun_apl}.
 
\begin{figure}[htb]
\sidecaption
  \includegraphics[width=.52\textwidth]{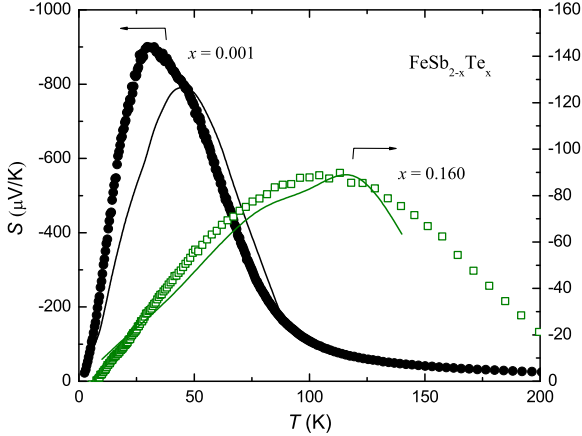}
\caption{Thermopower $S(T)$ for semiconducting  FeSb$_{1.999}$Te$_{0.001}$, and metallic FeSb$_{1.84}$Te$_{0.16}$. Solid lines denote results of calculations based on the measured Hall 
coefficient and additional enhancement factor (see text), presumably caused by many-body renormalizations \cite{sun_apl}. }
\label{fig:S}
\end{figure} 

The most remarkable transport phenomenon of FeSb$_2$ is its colossal negative thermopower $S$ ranging from 6 to $-$45 mV/K at $T$ $\approx$ 10 K \cite{bentien}. For a classical 
semiconductor, the possible maximum thermopower at a specific temperature is determined by the magnitude of the relevant energy gap, i.e., $S_{max}$ = $E_g$/2$eT_{max}$ \cite{goldsmid}. In 
view of the small magnitude of the transport gaps, this approach yields a thermopower of $\sim$1.5 mV/K at 10 K. By analyzing the temperature profile of the thermopower of FeSb$_2$ 
\cite{sun_apl,sun_PRB}, and comparing it with that of a classical homologue \cite{sun_apex}, we concluded that the thermopower of FeSb$_2$ can be {\it qualitatively} described  by the 
classical expression for electron diffusion. However, a {\it quantitative} description requires an enhancement factor 10$-$30. Surprisingly, such a large thermopower enhancement was also 
found to persist when crossing the SM transition induced by doping \cite{sun_apl}.  This is demonstrated in Fig.~\ref{fig:S}, which displays the thermopower of two typical samples, i.e.,  
FeSb$_{1.999}$Te$_{0.001}$ (semiconducting)  and  FeSb$_{1.84}$Te$_{0.16}$ (metallic). The solid lines plotted in this figure are based on the following classical formula for degenerate 
system \cite{sun_apl},
\begin{equation}
S(T)\,=\,\frac{\pi ^2}{3} \frac{k_{B}}{e} \frac{k_BT}{\epsilon _F},
\end{equation}
where the Fermi energy $\epsilon _F$ is determined by the carrier concentration $n$ and the effective mass $m^*$ of charge carriers, 
$\epsilon _F\,=\,\frac{h^2}{2m^*}(\frac{3n}{8\pi})^{2/3}$. 
By assuming $m^*$ = $m_0$, the free electron mass, and applying the one-band model where $n$ $=$ 1/$e$$|R_H|$, an enhancement factor of 18 or 32 had to be applied to Eq. 1 in order to 
reproduce the measured $S(T)$. This is in agreement with a corresponding factor 10$-30$ found in undoped FeSb$_2$ \cite{sun_PRB}.    

Our qualitative description of $S(T)$ with the aid of the classical expression strongly suggests a dominating electronic rather than phononic origin of the thermopower.  A non-negligible  
contribution due to the phonon drag effect may nevertheless exist, particularly for the semiconducting FeSb$_{2-x}$Te$_x$ samples with low carrier concentration and enhanced phonon thermal 
conductivity. We speculate that the large enhancement (by a factor of 10$-$30) of the electron-diffusion thermopower has its origin in substantial electron-electron correlations. 
In at least the metallic regime ($x$$>$0.005), this can be well captured by a largely renormalized value of $m^*$, as is supported by the enhanced electronic specific heat (to be shown in 
Fig.~\ref{fig:C}).  This resembles the enhanced thermopower observed in 4$f$/5$f$-based heavy-fermion metals \cite{behnia}. 
As far as the semiconducting FeSb$_{2-x}$Te$_x$ ($x$ $<$ 0.003) samples are concerned, the physical interpretation of the thermopower enhancement given above appears to be inadequate: here, 
the charge carrier system tends to be nondegenerate (or on the borderline between degenerate and nondegenerate \cite{sun_apl,sun_PRB}). In this case, a renormalized $m^*$ does not 
effectively enhance $S(T)$ in the same way as in metallic systems.
In fact,  
recent theoretical calculations for FeSb$_2$ yield an only qualitative description of the observed thermopower \cite{Kotliar}. A quantitative description of nondegenerate semiconducting 
systems involving strong electron-electron correlations remains challenging.   

\section{Optical properties}

\begin{figure}[htb]
\sidecaption
  \includegraphics[width=.5\textwidth]{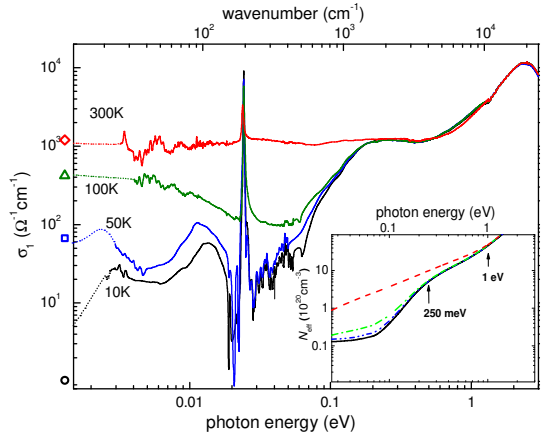}
\caption{Optical conductivity $\sigma _1 (\omega)$ of FeSb$_2$ with {\bf E$||$C} in the (110) plane. The symbols indicate the dc conductivity $\sigma _{dc}$ = $\sigma _1$ ($\omega$ = 0) at 
the respective temperatures. Inset: Energy dependence of the effective charge carrier density $N_{\rm eff}(\omega)$ \cite{herzog}.}
\label{fig:opti}
\end{figure}

The optical conductivity $\sigma _1 (\omega)$ of FeSb$_2$ was first reported by Perucchi $et\,al$. \cite{perucchi} and subsequently by Herzog $et\,al$. \cite{herzog}. Some of the recent data 
adapted from Ref. \cite{herzog} are shown in Fig.~\ref{fig:opti}. In the far-infrared range, $\sigma _1 (\omega)$ exhibits a pronounced temperature dependence. While at $T$ $<$ 100 K, there 
are clear gap features with diminishing Drude component, upon warming the system up to room temperature, the gaps are entirely filled up by thermal excitations, resulting in a flat $\sigma 
_1(\omega)$.  Analyzing the spectra
with a fundamental absorption across the gap of parabolic bands yields not only a direct gap at 130 meV, but also two indirect gaps at 6 meV and 31 meV \cite{herzog}. The observed direct gap 
is roughly half of that (0.2$-0.3$ eV) estimated by LDA \cite{bentien2,madsen, LDA}.  The indirect gaps, on the other hand,  correspond well to the transport gaps inferred from resistivity 
measurements (cf. Table 1). In stark contrast to classical band semiconductors, where the spectral weight lost by the gap formation is shifted to just above the gap energy, a remarkable 
observation in FeSb$_2$ is the large rearrangement of the spectral weight up to much higher energies. The two energies for FeSb$_2$ at which the lost spectral weight is recovered, i.e., one 
at 0.25 eV for $T$ $\leq$ 100 K and another one at 1 eV for $T$ $\leq$ 300 K (inset of Fig.~\ref{fig:opti}) largely exceed those of both the indirect and the direct gap. Such a strong 
spectral redistribution on a large (atomic) energy scale has also been observed for, e.g.,  FeSi \cite{opti_FeSi} and is considered characteristic for strongly correlated semiconductors.

\section{Specific heat} 

\begin{figure}[htb]
\sidecaption
  \includegraphics[width=.5\textwidth]{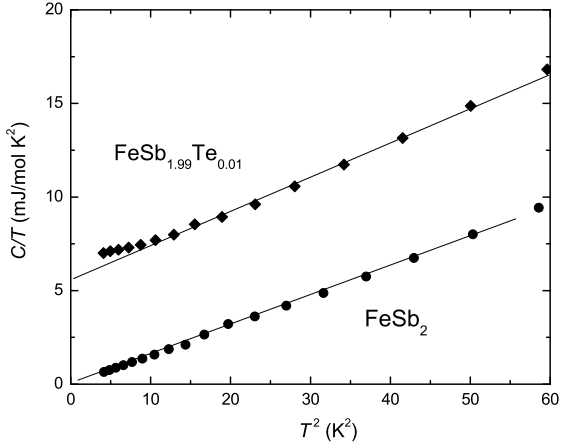}
\caption{Specific heat divided by temperature $C/T$ as a function of $T^2$ for FeSb$_2$ and FeSb$_{1.99}$Te$_{0.01}$. The electronic specific-heat coefficient $\gamma$ is read off as the 
intercept at $T$ = 0.}
\label{fig:C}
\end{figure}

Fig.~\ref{fig:C} shows the specific heat divided by temperature, $C/T$, as a function of $T^2$ for nominally pure FeSb$_2$ as well as  for FeSb$_{1.99}$Te$_{0.01}$. The low-temperature 
upturn observed for the latter sample may be due to either slight impurities or doping-induced short range cooperative magnetism,  which parallels the Curie tail at low temperatures (cf. 
Fig.~\ref{fig:chi}). The extrapolated electronic specific-heat coefficient $\gamma$ for the FeSb$_2$ sample shown in Fig.~\ref{fig:C} is practically zero. Depending on the differing purities 
of various specimens investigated, some nominally pure FeSb$_2$ samples show finite $\gamma$ values, however, always smaller than 0.25 mJ/mol K$^2$.  
In line with the metallic conduction observed in the Te-doped systems, the introduced free carriers contribute a term $\gamma T$ to the specific heat. For the metallic sample 
FeSb$_{1.99}$Te$_{0.01}$, $\gamma$ is as large as 6 mJ/mol K$^2$. Based on the carrier concentration $n$\,=\,$1/e|R_H|$\,=\,1.7 $\times$ 10$^{20}$ cm$^3$ derived from Hall coefficient (cf. 
Fig.~\ref{fig:hall}) and from the free-electron expression $\gamma$ = $\pi ^2$/3 $k_B^2$$N(E_F)$, an effective quasiparticle mass $m^*$ of $\sim$18 times the bare electron mass $m_0$ is 
estimated for FeSb$_{1.99}$Te$_{0.01}$. The enhanced $m^*$ ($=$ 10$-$20 $m_0$) appears to be almost doping independent in the whole concentration range $x$ $<$ 0.16 investigated by us 
\cite{unpublish}. 
Interestingly, the estimated  $m^*$ for Te-doped FeSb$_2$ is comparable to $m^*$ $\approx$ 14 $m_0$ as estimated for Al-doped FeSi based on specific heat measurements \cite{FeSiAl}. 

This enhanced electronic specific heat observed for Te-doped FeSb$_2$ is compatible with the doping induced, $T$-independent Pauli paramagnetism discussed in section 2.  From the observed 
$\gamma$ value (6 mJ/mol K$^2$) for FeSb$_{1.99}$Te$_{0.01}$, we estimate $N(\varepsilon _F)$ = 0.96 $\times$ 10$^{43}$ states$/$J mol, in good agreement with $N(\varepsilon _F)$ = 1.39 
$\times$ 10$^{43}$ states$/$J mol obtained from the doping induced Pauli paramagnetism. Using this latter value, one find $m^*$ $\approx$ 26 $m_0$, close to 18 $m_0$ as estimated from the 
specific heat. The enhanced quasiparticle mass, which is not captured by the LDA band structure calculations \cite{LDA}, provides a clue to understanding the observed enhancement of the 
thermopower by a similar factor \cite{sun_apl}.

\section{Discussion and summary}

 The unusual gap opening of FeSb$_2$ was also detected by other measurements, i.e., of the NQR \cite{nqr1,nqr2} and the M\"{o}ssbauer \cite{moss_FeSb2}. The formation of a small gap with 
activation energy $\Delta$ = 38.4 meV was seen in the spin-lattice relaxation rate, 1/$T_1$($T$) \cite{nqr1,nqr2}. The corresponding gap 2$\Delta$ (as expected for an intrinsic semiconductor 
with symmetric valence and conduction bands) is close to what was estimated from the static magnetic susceptibility and appears to reflect the direct gap of FeSb$_2$ seen in the optical 
measurements (cf. Table 1). On the other hand, the energy gap derived from the temperature-dependent quadrupole splitting of the M\"{o}ssbauer effect is 33 meV, in good agreement with both 
the transport gap ($E_{g2}$) and the indirect gap seen in the optical conductivity data (cf. Table 1).   
It is worthwhile to note that different excitation energies for charge and spin channels, 
roughly corresponding to the indirect and direct gaps, have also been observed for FeSi \cite{sales}. 

Though LDA band structure calculation is apt to predict the formation of an energy gap in FeSb$_2$, it cannot capture the unusual physical properties observed in this material. These include 
a thermally activated and enhanced Pauli paramagnetism, a strong rearrangement of optical spectral weight, a colossal thermopower as well as an enhanced quarsiparticle mass $m^*$. All these 
phenomena distinguish FeSb$_2$ from classical band insulators. Furthermore, the effects of many body renormalization inferred from these measurements seem consistent: like the $m^*$ that is 
10$-$30 times enhanced as indicated by the specific heat and magnetic susceptibility measurements, the thermopower exhibits an enhancement by a similar factor. 

Ferromagnetism induced by doping in FeSb$_2$ \cite{Hu_Te, Hu_Co} as well as in FeSi \cite{LDA_FeSi,Yeo} is another intriguing fundamental issue relevant to electron correlations in 
low-carrier systems. The induced ferrommagnetism has been so far discussed in either TM's nearly ferromagnetic semiconductor \cite{LDA_FeSi} or the KI scenario \cite{Yeo} and is also 
understandable in the recent CBI picture \cite{CBI}. The nature of the magnetically ordered state is yet to be further clarified. Similarly, the heavy-fermion like metallic state induced by 
doping is far from being understood. Different to the Kondo scenario that is valid for $f$-based KI's, in a CBI, the Coulomb correlations lead to band narrowing and a corresponding mass 
renormalization near the Fermi level. However, it remains to be shown why the CBI approach yields only $m^*$ $\sim$ 2$m_0$ (cf. \cite{CBI}), much smaller than the values estimated from the 
thermodynamic results. Like in FeSi (or its doped systems), where magnetic field can induce anomalous Hall conductance 
\cite{Manyala1} and unconventional magnetoresistance \cite{Manyala2}, similar effects are argued to exist in FeSb$_2$ as well \cite{Hu_MR}. These phenomena are of particular technologically 
interest and are calling for further investigations.    

In summary, we have described various physical properties of the narrow-gap semiconductor FeSb$_2$. The anomalous temperature / energy dependences observed in transport, optical and 
thermodymamic properties cannot be understood in the framework of conventional band theory.  Electron-electron correlations are believed to be at the origin of the strongly renormalized 
electronic properties as found in various measurements.  By slightly doping Te (less than 0.5\%), FeSb$_2$ can be made metallic and shows behaviors reminiscent of $f$-based heavy fermion 
metals, in particular a renormalized quasiparticle mass as reflected by large low-$T$ values of the electronic specific heat and diffusion thermopower.

\begin{acknowledgement}
We thank N. Oeschler, J. Simon, Y. Sun, A. Bentien, J. Sichelschmidt, and M. Baenitz for fruitful discussions. 

\end{acknowledgement}


\begin{thebibliography}{10}

\bibitem{Jacca} V. Jaccarino, G.K. Wertheim, J.H. Wernick, L.R. Walker, and S. Arajs, Phys. Rev. \textbf{160}, 476 (1967).
\bibitem{review_KI} P.S. Riseborough, Adv. Phys. \textbf{49}, 257 (2000).
\bibitem{fisk} G. Aeppli and Z. Fisk, Comments Condens. Matter Phys.  \textbf{16}, 155 (1992).
\bibitem{FeVAl} Y. Nishino, M. Kato, S. Asano, K. Soda, M. Hayasaki, and U. Mizutani, Phys. Rev. Lett. \textbf{79}, 1909 (1997).
\bibitem{taka} Y. Hadano, S. Narazu, M.A. Avila, T. Onimaru, and T. Takabatake, J. Phys. Soc. Jpn. \textbf{78}, 013702 (2009).
\bibitem{petrovic1} C. Petrovic, Y. Lee, T. Vogt, N.Dj. Lazarov, S.L. Bud'ko, and P.C. Canfield, Phys. Rev. B {\bf 72}, 045103 (2005). 
\bibitem{petrovic2} C. Petrovic, J.W. Kim, S.L. Bud'ko, A.I. Goldman, P.C. Canfield, W. Choe, and G.J. Miller,  Phys. Rev. B {\bf 67}, 155205 (2003).
\bibitem{bentien2}  A. Bentien, G. K. H. Madsen, S. Johnsen, and B. B. Iversen, Phys. Rev. B {\bf 74}, 205105 (2006).

\bibitem{SCR} Y. Takahashi and T. Moriya, J. Phys. Soc. Jpn. \textbf{46}, 1451 (1979).
\bibitem{kunes} J. Kune\v{s} and V.I. Anisimov, Phys. Rev. B \textbf{78}, 033109 (2008).
\bibitem{Sentef} M. Sentef, J. Kune\v{s}, P. Werner, and A.P. Kampf, Phys. Rev. B \textbf{80}, 155116 (2009).
\bibitem{CBI} V.V. Mazurenko, A.O. Shorikov, A.V. Lukoyanov, K. Kharlov, E. Gorelov, A.I. Lichtenstein, and V.I. Anisimov, Phys. Rev. B \textbf{81}, 125131 (2010).
\bibitem{dmft} G. Kotliar, and D. Vollhardt, Phys. Today \textbf{57}, 53 (2004).
\bibitem{Kotliar} J. M. Tomczak, K. Haule, T. Miyake, A. Georges, and G. Kotliar, Phys. Rev. B {\bf 82}, 085104 (2010).

\bibitem{klein} M. Klein, D. Zur, D. Menzel, J. Schoenes, K. Doll, J. R\"{o}der, and F. Reinert, Phys. Rev. Lett. \textbf{101}, 046406 (2008).
\bibitem{marca} F. Hulliger, Nature \textbf{198}, 1081 (1963).
\bibitem{Fan} A.K.L. Fan, G. H. Rosenthal, H.L.Mckinzie and A. Wold, J. Solid State Chem. \textbf{5}, 136 (1972).
\bibitem{silve} C.E.T. Goncalves da Silva, Solid State Commun. \textbf{33}, 63 (1980).

\bibitem{madsen} G.K.H. Madsen, A. Bentien, S. Johnsen, and B.B. Iversen, in Proceedings of the 25th International Conference on thermoelectrics (IEEE, New York,2006), pp. 579-581.
\bibitem{LDA} A.V. Lukoyanov, V.V. Mazurenko, V.I. Anisimov, M. Sigrist, and T.M. Rice, Eur. Phys. J. B {\bf 53}, 205 (2006).
\bibitem{LDA_FeSi} V.I. Anisimov, S.Y. Ezhov, I.S. Elfimov, I.V. Solovyev, and T.M. Rice, Phys. Rev. Lett. \textbf{76}, 1735 (1996). 

\bibitem{Yeo} S. Yeo, S. Nakatsuji, A.D. Bianchi, P. Schlottmann, Z. Fisk, L. Balicas, P.A. Stampe, and R.J. Kennedy, Phys. Rev. Lett. \textbf{91}, 046401 (2003).
\bibitem{Hu_Te} R. Hu, V.F. Mitrovi\'{c}, and C. Petrovic, Phys. Rev. B \textbf{79}, 064510 (2009).
\bibitem{Hu_Co} R. Hu, V.F. Mitrovi\'{c}, and C. Petrovic, Phys. Rev. B \textbf{74}, 195130 (2006).
\bibitem{FeSiAl} J.F. DiTusa, K. Friemelt, E. Bucher, G. Aeppli and A.P. Ramirez, Phys. Rev. B \textbf{58}, 10288 (1998). 
\bibitem{sun_apl} P. Sun, M. S\o ndergaard, Y. Sun, S. Johnsen, B.B. Iversen, and F. Steglich, Appl. Phys. Lett. \textbf{98} (2011), to appear. 
\bibitem{bentien} A. Bentien, S. Johnsen, G. K. H. Madsen, B. B. Iversen, and F. Steglich, Europhys. Lett.  {\bf 80}, 17008 (2007).

\bibitem{sun_PRB} P. Sun, N. Oeschler, S. Johnsen, B.B. Iversen, and F. Stgelich, Phys. Rev. B \textbf{79}, 153308 (2009). 
\bibitem{sun_apex} P. Sun, N. Oeschler, S. Johnsen, B.B. Iversen, and F. Stgelich, Appl. Phys. Express \textbf{2}, 091102 (2009).

\bibitem{neutron_FeSb2} H. Holseth, and A. Kjekshus, Acta Chem. Scand. \textbf{24}, 3309 (1970).
\bibitem{moss_FeSb2} J. Steger, and E. Kostiner, J. Solid State Chem. \textbf{5}, 131 (1972).
\bibitem{goodenough} J.B. Goodenough, J. Solid State Chem. \textbf{5}, 144 (1972).
\bibitem{koyama} T. Koyama, H. Nakamura, T. Kohara, and Y. Takahashi, J. Phys. Soc. Jpn. \textbf{79}, 093704 (2010).
\bibitem{nqr1} T. Koyama, Y. Fukui, Y. Muro, T. Nagao, H. Nakamura, and T. Kohara, Phys. Rev. B \textbf{76}, 073203 (2007).
\bibitem{mandrus} D. Mandrus, J. L. Sarrao, A. Migliori, J.D. Thompson, and Z. Fisk, Phys. Rev. B \textbf{51}, 4763 (1995).

\bibitem{decompose} While Fan et\,al. \cite{Fan} reported that FeSb$_2$ under vacuum starts to decompose into FeSb and Sb at 573 K, different authors found that bulk FeSb$_2$ is chemically 
stable up to around 1000 K. See, for example, F. Gr\o nvold et\,al., J. Chem. Thermodynamics \textbf{9}, 773 (1977).

\bibitem{sun_dalton} P. Sun, N. Oeschler, S. Johnsen, B.B. Iversen, and F. Stgelich, Dalton Trans., \textbf{39}, 1012 (2010).
\bibitem{goldsmid} H.J. Goldsmid and J.W. Sharp, J. Electron. Mater. \textbf{28}, 869 (1999).
\bibitem{behnia}  K. Behnia, D. Jaccard, and J. Flouquet, J. Phys.: Condens. Matter {\bf 16}, 5187 (2004).
\bibitem{perucchi} A. Perucchi, L. Degiorgi, R. Hu, C. Petrovic and V.F. Mitrovi\'{c}, Eur. Phys. J. B \textbf{54}, 175 (2006).
\bibitem{herzog} A. Herzog, M. Marutzky, J. Sichelschmidt, F. Steglich, S. Kimura, S. Johnsen, and B.B. Iversen, Phys. Rev. B \textbf{82}, 245205 (2010).
\bibitem{opti_FeSi} Z. Schlesinger, Z. Fisk, H.-T. Zhang, M.B. Maple, J.F. DiTusa, and G. Aeppli, Phys. Rev. Lett. \textbf{71}, 1748 (1993).
\bibitem{unpublish} P. Sun, M. S\o ndergaard, B.B. Iversen, and F. Steglich, unpublished. 
\bibitem{nqr2} A.A. Gippius, K.S. Okhotnikov, M. Baenitz, and A.V. Shevelkov, Solid State Phenom. \textbf{152-153}, 287{2009}.
\bibitem{sales} B.C. Sales, E.C. Jones, B.C. Chakoumakos, J.A. Fernandez-Baca, H.E. Harmon, J.W. Sharp, and E.H. Volckmann,  Phys. Rev. B \textbf{50}, 8207 (1994).

\bibitem{Manyala1} N. Manyala et\,al., Nature Mater. \textbf{3}, 255 (2004).
\bibitem{Manyala2} N. Manyala, Y. Sidis, J.F. DiTusa, G. Aeppli, D.P. Young, and Z. Fisk, Nature  \textbf{404}, 581 (2000).
\bibitem{Hu_MR} R. Hu et\,al., Phys. Rev. B \textbf{77}, 085212 (2008).
 

\end{thebibliography}
\end{document}